\documentclass{JINST}

\title{Daya Bay Antineutrino Detector Gas System}

\author{H.~R.~Band$^a$\thanks{Corresponding
author.}, J.~J.~Cherwinka$^b$,~M-C.~Chu$^c$, K.~M.~Heeger$^a$, M.~W.~Kwok$^c$, K.~Shih$^c$, T.~Wise$^a$,~and Q.~Xiao $^b$\\
\llap{$^a$}Department of Physics, University of Wisconsin - Madison,\\
  1150 University Avenue, Madison, WI 53706, U.~S.~A\\
\llap{$^b$}Physical Sciences Laboratory, University of Wisconsin - Madison,\\
  3725 Schneider Drive, Stoughton, WI 53589, U.~S.~A\\
  \llap{$^c$}Chinese University of Hong Kong,\\
 Hong Kong\\
  E-mail: \email{hrb@slac.stanford.edu}}

\abstract{The Daya Bay Antineutrino Detector gas system is designed to protect the liquid scintillator targets of the antineutrino detectors against degradation and contamination from exposure to ambient laboratory air. The gas system is also used to monitor the leak tightness of the antineutrino detector assembly. The cover gas system constantly flushes the gas volumes above the liquid scintillator with dry nitrogen to minimize oxidation of the scintillator over the five year lifetime of the experiment. This constant flush also  prevents the infiltration of radon or other contaminants into these detecting liquids keeping the internal backgrounds low. 
Since the Daya Bay antineutrino detectors are immersed in the large water pools of the muon veto system, other
gas volumes are needed to protect vital detector cables or gas lines. These volumes are also purged with dry gas. Return gas is monitored for oxygen content and humidity to provide early warning of 
potentially damaging leaks. The design and performance of the Daya Bay Antineutrino Detector gas system is described.}


\begin{document}

\section{Introduction}

        The neutrino mixing angle $\theta_{13}$ was the last unmeasured angle in the neutrino mixing matrix~\cite{bib1,bib2}
before recent measurements by Daya Bay~\cite{DB}, RENO ~\cite{RENO} and Double Chooz~\cite{Chooz}. The mixing angle $\theta_{13}$ is a fundamental parameter of the new Standard Model and accurate knowledge of $\theta_{13}$ is
needed to plan future experiments designed to measure the neutrino mass hierarchy and CP
asymmetry. Daya Bay is using electron-type antineutrinos from six high power (2.9 GW$_{th}$)
commercial nuclear reactors to measure a deviation from the expected $1/r^2$ behavior in the
number of antineutrino interactions observed as a function of distance from the nuclear reactor
cores. Interpreting the observed deficit as evidence of neutrino oscillation,  $\sin^2 2\theta_{13}$, was measured as $0.089\pm
0.010~\rm{(stat}) ~\pm0.005 (\rm{syst})$ ~\cite{DB} within the framework of 3-neutrino mixing.

Antineutrinos are detected via the inverse beta-decay reaction (IBD), $\overline{\nu}{_e} + p \rightarrow e^+~+~n$ . The
positron annihilates in the liquid scintillator producing a prompt energy pulse. The
neutron thermalizes before being absorbed by a gadolinium or hydrogen nucleus and produces an
energy pulse delayed by typically 30 $\mu$sec. This characteristic time-correlated energy signal allows the antineutrino
signals to be cleanly separated from the copious single-energy depositions generated by radioactive
backgrounds.

The Daya Bay antineutrino detectors (ADs) were carefully constructed in a class 10,000 clean room from low-radioactivity materials to reduce internal detector backgrounds. Equal care was taken with the liquid scintillator base and additives. During data taking at Daya Bay the ADs are operated  in water pools which strongly attenuate 
ambient radiation from the underground granite experimental halls and also serve as cosmic ray vetoes.  To maintain this low radioactivity system it is necessary to prevent the infiltration of radon gas, prevalent in underground environments, and other airborne contaminants into the AD detecting liquids. Radon-222 decays with a half-life of 3.8 days into a long decay chain of radioactive isotopes, each of which generate signals in the liquid scintillator. Since radon can permeate any of the numerous o-rings in the detector, a design was chosen which flushes all of the gas volumes above the detector liquids with an inert gas. Boil-off nitrogen is used as the cover gas. This design also reduces the risk of oxygen exposure, which has been shown to degrade the performance of the scintillating agents used in the AD detecting liquids ~\cite{tdr}.
        
The gas system was expanded to purge other gas volumes on top of the detector that protect vital detector cabling or gas lines. Returns from all of these gas circuits are continuously monitored for relative humidity to provide an early warning of water leaks as the antineutrino detectors are immersed beneath 2.5 meter of water. The return gases are also periodically checked for radon or oxygen to ensure that the gas system is operating properly. 

\subsection{Daya Bay Reactor Neutrino Experiment}

\begin{figure}[t]\hfil
\includegraphics[clip=true, trim=50mm  35mm 60mm 75mm,width=5.5in]{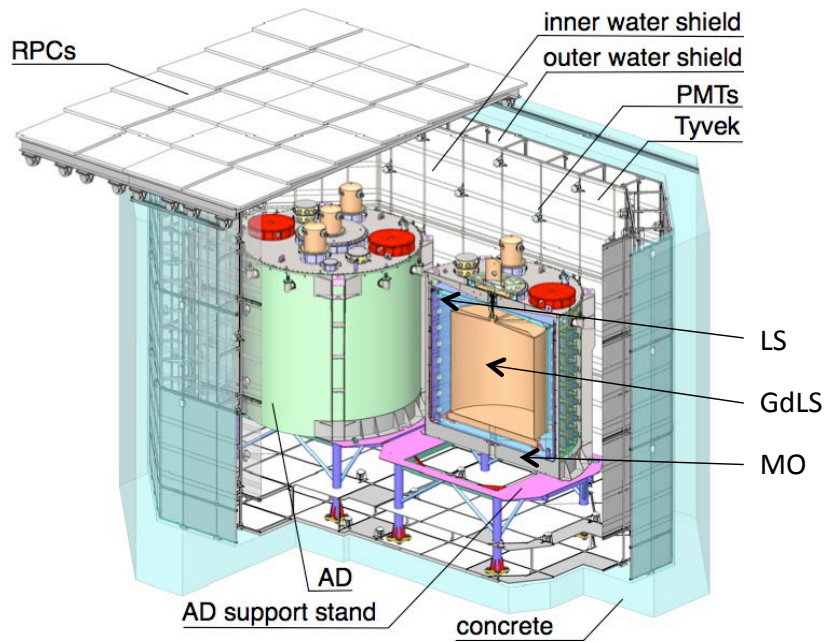}\hfil%
\caption{Two antineutrino detectors are shown in a water pool instrumented with PMTs. 
Each AD contains 3 zones: a gadolinium loaded liquid scintillator inner target
(20 tons) inside the inner acrylic vessel (brown), a liquid scintillator gamma catcher (21 tons) contained by the outer acrylic vessel (blue), and a mineral oil buffer (37 tons) inside the stainless steel outer vessel (grey).
The water pool is split into inner and outer zones by layers of Tyvek.}\label{fig:EH1}
\end{figure}

The Daya Bay experiment consists of 3 underground experimental halls located 360-1900 m  from
the reactor cores. Each of the two near detector halls contains two antineutrino detectors in a water pool. The far experimental hall contains four ADs in a  larger water pool.
Each AD consists of a Stainless Steel Vessel (SSV) containing 3 detector zones filled with
different liquids as shown in Fig.~\ref{fig:EH1}. Two nested acrylic cylinders separate the 3 zones. The
innermost target zone contains 20 tons of gadolinium-doped liquid scintillator (GdLS) inside an
Inner Acrylic Vessel (IAV). This zone is surrounded by 21 tons of liquid scintillator (LS)
gamma catcher contained by an Outer Acrylic Vessel (OAV). The energy from IBD interactions
is observed by 192 photomultipliers (PMTs) arranged in a cylindrical shell. An outer zone of
liquid contains 37 tons of mineral oil (MO) that shield the inner zones from the radioactivity of
the glass PMTs and other background sources. Reflectors above and below the LS volume
improve the energy response of the detector and  make it more uniform. Three automated calibration units (ACU) above the SSV lid
allow remote deployment of radioactive sources or LED flashers into the GdLS or LS liquid
volumes. Weekly calibration data runs are interspersed with normal data to accurately measure
and track the energy response of each AD.

The ADs sit in water pools which shield the detectors  in all directions
from ambient radioactivity. The water pools are instrumented with PMTs arranged in optically
separated inner (IWS) and outer water (OWS) pool zones to detect muons which may introduce  spallation neutrons or other cosmogenic backgrounds into the ADs. Additional muon detection is provided by four layers of resistive plate chambers (RPCs) which can be rolled over the water pool.

\section{Detector Gas System}

\begin{figure}\hfil
\includegraphics[clip=true, trim=0mm  72mm 10mm 55mm,width=5.in]{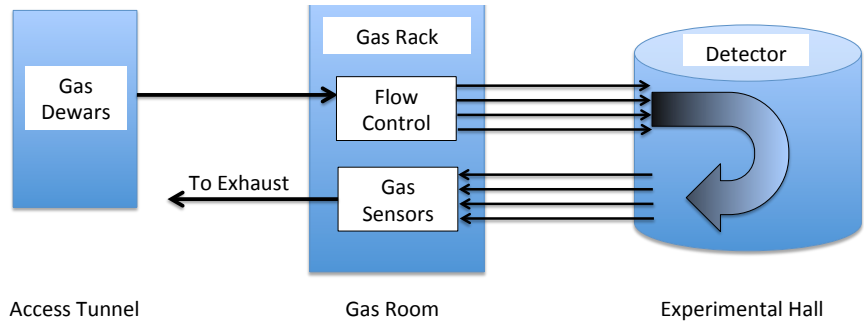}\hfil
\caption{Nitrogen from dewars located in the access tunnels is routed to a gas distribution rack near the experimental hall. The gas flow for each AD  is split into 4 gas circuits  each of which is sent to the AD. Return gas is monitored for humidity and oxygen.}\label{fig:Overview}
\end{figure}

        The flow of gas through the AD gas system is shown schematically in Fig.~\ref{fig:Overview}. Boil-off from liquid nitrogen
dewars stored in the Daya Bay access tunnels near each experimental hall is pressure regulated 
and sent to a gas room adjacent to the Experimental Hall (EH). The  AD gas rack inside the gas room 
splits the gas  into separate circuits for each AD. Each AD flow is further divided into four gas
circuits controlled by manual flow-meters to $\sim~1-2$  volume exchanges per day. The gas is then routed to
the AD to purge the specified gas volumes. Return gas is collected and sent back to the gas rack.
The returning gas humidity is measured before the gas bubbles through a mineral oil  exit bubbler and into
the experimental hall exhaust vent.  The bubblers keep the gas volume pressure above atmosphere, reducing back infiltration and provide a visual confirmation of the gas flow.  The oxygen content of the supply gas or one of the gas returns can be additionally measured. 
Although the gas system in each hall is operated manually, the system is continuously  monitored by  electronics collecting
pressure, humidity, and oxygen content information interfaced to the Daya Bay monitoring
software ~\cite{tdr,DCS}, which records this data into an online database and makes it available to shift
personnel.

The boil-off from the commercial grade liquid nitrogen dewars available in China meets our gas
quality needs. The oxygen contamination has been measured to be less than 5 ppm. The relative
humidity was measured as < 0.5\% (at room temperature). Radioactivity levels were below the sensitivity of
the measurement apparatus ($\approx 5$ mBq per m$^3$).

\subsection{Cover Gas}

\begin{figure}[b]\hfil  
\includegraphics[clip=true, trim=10  58mm 5mm 45mm,width=5.8in]{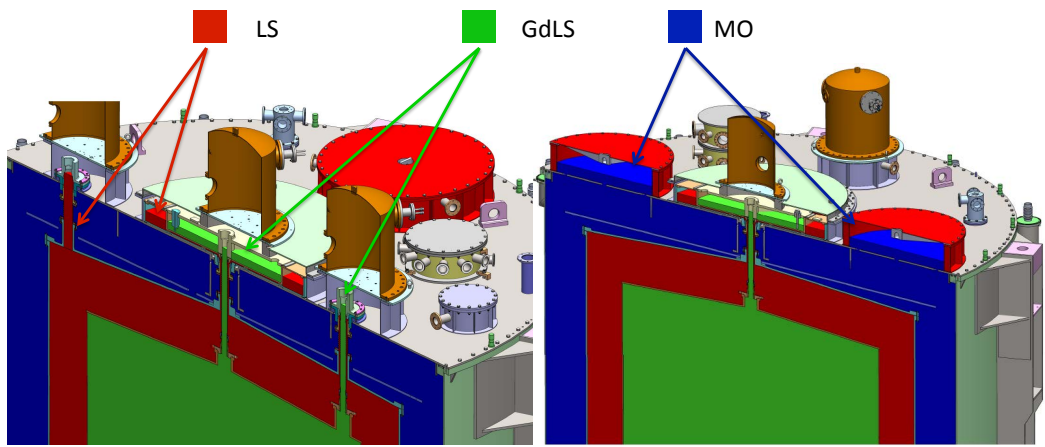}\hfil%
\caption{GdLS (green), LS (red), and MO (blue) overflow tanks and calibration tubes. }\label{fig:CG}
\end{figure}

        The primary function of the gas system is to provide a clean, inert gas blanket over the exposed liquids in the overflow tanks shown in Fig.~\ref{fig:CG}.  The overflow tanks are located above the central target volumes and allow for the thermal expansion of the liquids as well as the precise measurement of the target mass ~\cite{Mass}. A central circular GdLS overflow tank
rests inside a larger diameter LS overflow tank with both tanks sharing a common gas volume.
Two circular MO overflow tanks flank the center. Small areas of liquid are exposed in each of the three calibration tubes.

\begin{figure}\hfil
\includegraphics[clip=true, trim=5  15mm -5mm 15mm,width=4.5in]{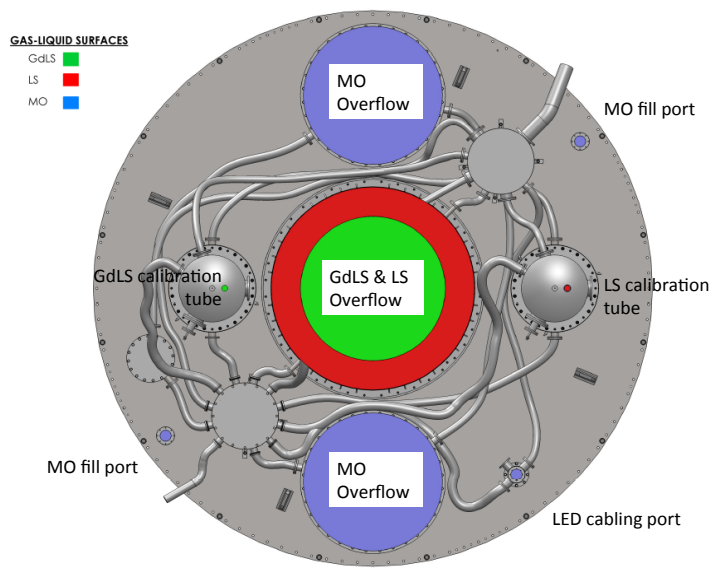}\hfil%
\caption{Top view of an AD showing the cover gas/liquid interfaces.}\label{fig:CGliquid}
\end{figure}

The gas/liquid surface areas shown in Fig.~\ref{fig:CGliquid} are dominated by the overflow tanks and are approximately
1.0 m$^2$ (GdLS), 1.1 m$^2$ (LS), and 2.1 m$^2$ (MO). The gas volume in the MO overflow tanks is
about 300 liters assuming the tank is half full. The gas volumes above each off-axis port are split 
between the base ($\approx 80$~liters) and ACU above ($\approx 210$~liters). 
The gas volume above the central overflow tanks is  $\approx 380$~ liters.  The total gas volume of the cover 
gas system is $\approx 1500$~liters. Flushing two volume exchanges per day requires a flow rate of 2.1 liters per minute.
The flows were typically set higher (4-5 lpm) during the first 2-4 weeks of operation to purge the system more quickly.

A major design goal was to ensure that differential pressures between the various gas volumes could not exceed more than a few cm of water column.
Otherwise, different pressure heads could pump liquids out of the calibration tubes or overfill the overflow tanks. This is accomplished by connecting
all cover gas supply and return lines to common manifolds on top of the SSV lid, limiting pressure differences in the final system to 1 mm. 
During many installation and leak checking activities some gas volumes are at atmosphere while gas is flowing through the remaining volumes. 
Restricting the absolute pressure by bubbling the return gas through $\sim2$ cm of oil limits the size of possible pressure transients during 
installation to $< 2$ cm.

\begin{figure}\hfil
\includegraphics[clip=true, trim=15  53mm 0mm 60mm,width=6.0in]{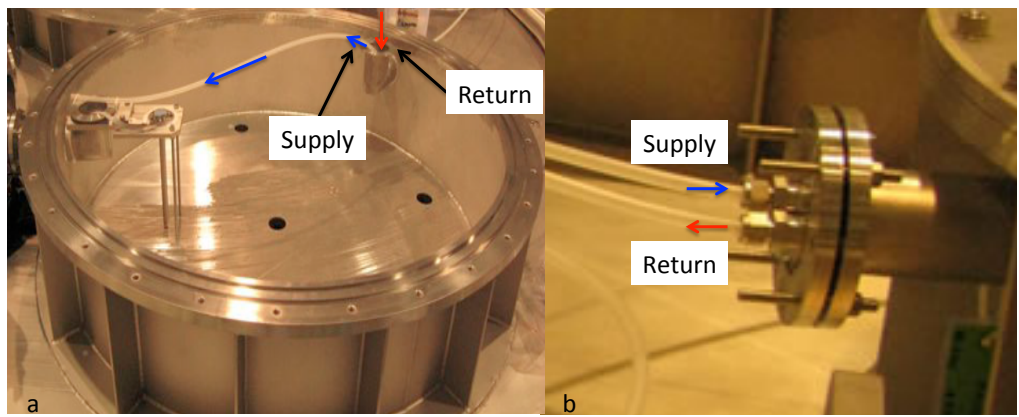}\hfil%
\caption{(a) Gas flow inside the MO overflow tank. The supply line enters  the volume on the right and delivers fresh gas on the left. Return gas is collected by the return line inside the right side elbow. (b) Typical gas feed through flange isolates the cover gas volume on the right from the gas volume inside bellows protecting the AD gas lines.}\label{fig:MO}
\end{figure}

To reduce radon infiltration, nitrogen from the gas rack is routed through a 12 mm copper line in 
the cabling trenches before transitioning at the edge of the water pool to a 3/4 inch polyethylene line 
which runs inside a stainless steel
bellows protecting the gas lines from the water pool. Inside the gas distribution
box on the detector lid the gas flow is split into 10 gas supply/return pairs. Each pair provides fresh gas to the
overflow tanks, ACUs, and special cabling and monitoring ports. The cover gas volumes shown in Fig.~\ref{fig:CGliquid} are
separated from other gas volumes on the AD lid by electrical and gas feed-through flanges.
The gas distribution inside each of the individual cover gas volumes is usually simple. The
supply line terminates on the far side of the volume being purged with the return line collecting gas at the
gas flange side. Fig.~\ref{fig:MO}a shows typical gas connections inside a MO overflow tank. 
Fig.~\ref{fig:MO}b shows a gas feed-through flange without the protecting gas dry pipe bellows.
        
\begin{figure}\hfil
\includegraphics[clip=true, trim=15  10mm 5mm 20mm,width=6.0in]{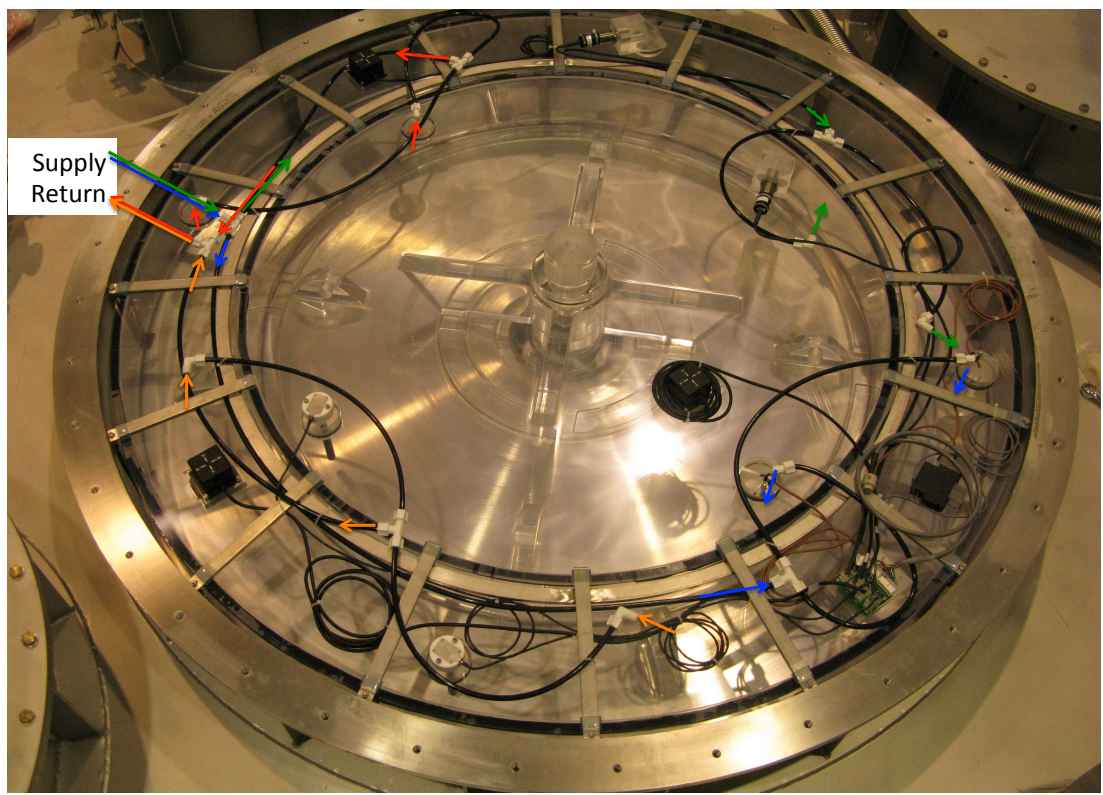}\hfil%
\caption{ Central overflow tanks with gas distribution tubing (black) and monitoring sensors. The circular GdLS tank is in the middle. The annular LS tank is on the outside. Gas enters and returns on the upper left. The supply gas is split into two legs. One leg (blue) is directed underneath the acrylic overflow tank lids 
on the right and purges the space above the liquids. The other leg (green) purges the space above the acrylic lid.  Return gas is collected above the acrylic lid (orange) or beneath the acrylic lid (red). }\label{fig:central}
\end{figure}

The gas flow in the central overflow volume (Fig.~\ref{fig:central}) is more complicated since the overflow tanks are
covered by acrylic lids with gas above and below the acrylic. The input gas flow is split into four. Two lines penetrate the acrylic lids above the GdLS and LS liquids. The other two lines flush the space above the acrylic lid.
The return gas is collected in a similar manner.
Supply and return lines are connected to opposite sides of the overflow tank gas volume to promote cross flow.
  
 The ACUs containing the radioactive sources are separately flushed in parallel with the gas volumes containing the calibration tubes (the central calibration tube is in the central overflow tanks). A 25 mm hole connects the ACU and calibration tube volumes. No attempt is made to direct gas flow through this hole but some gas sharing between these volumes is expected.

\subsection{Electrical Dry Pipe and Bellows}

\begin{figure}\hfil
\includegraphics[clip=true, trim=15mm  38mm -15mm 25mm,width=6.5in]{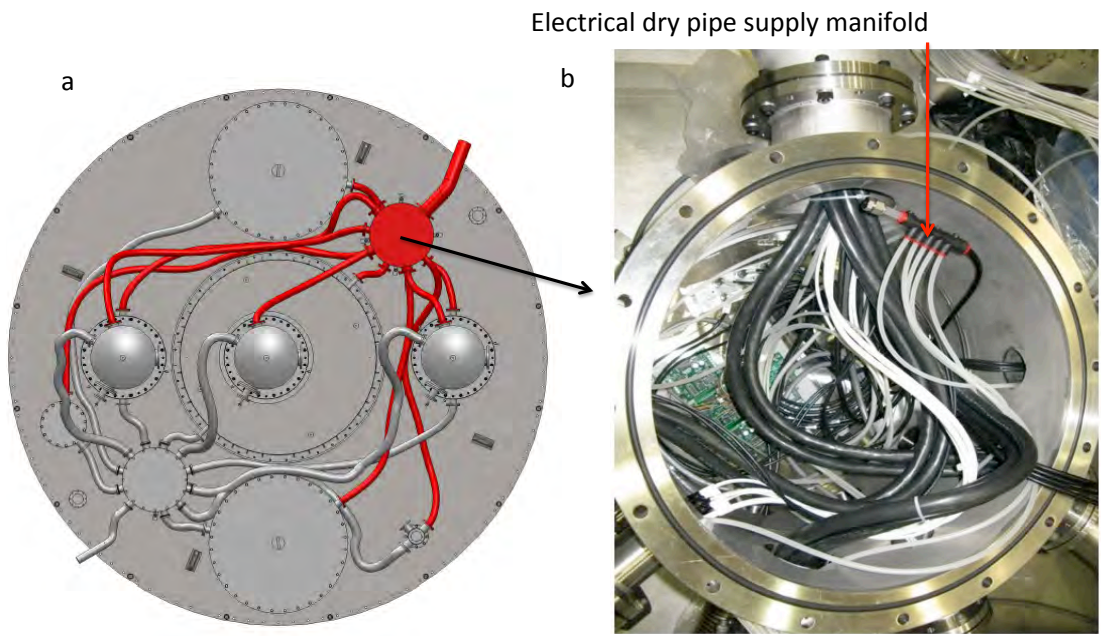}
\caption{(a) Common gas volume
(red) of the electrical distribution box and associated bellows. (b) Electrical distribution box with  dry pipe supply manifold and tubing inside.  }\label{fig:electrical}
\end{figure}

        The Daya Bay detector gas system also provides  a dry nitrogen atmosphere for the electrical connections and cabling inside and to the antineutrino detectors.  Electrical signals from monitoring sensors and calibration control units on the AD lid are gathered together in an electrical
distribution box (Fig.~\ref{fig:electrical}b) before being routed to the surface though a large vacuum bellows. The
electrical lines run through bellows on the lid which are isolated from the cover gas volumes described previously.
The on-lid bellows, distribution box, and long dry pipe to the surface comprise a single gas
volume shown in Fig.~\ref{fig:electrical}a. This volume is purged by a single 3/8  inch 
line which is split into 10 smaller lines. These
lines are run to the end of each of the on-lid bellows. The return gas flows back through the on-lid
bellows, through the distribution box, and through the electrical dry pipe to the surface.

\begin{figure}\hfil
\includegraphics[clip=true, trim=15mm 55mm 0mm 75mm,width=6.5in]{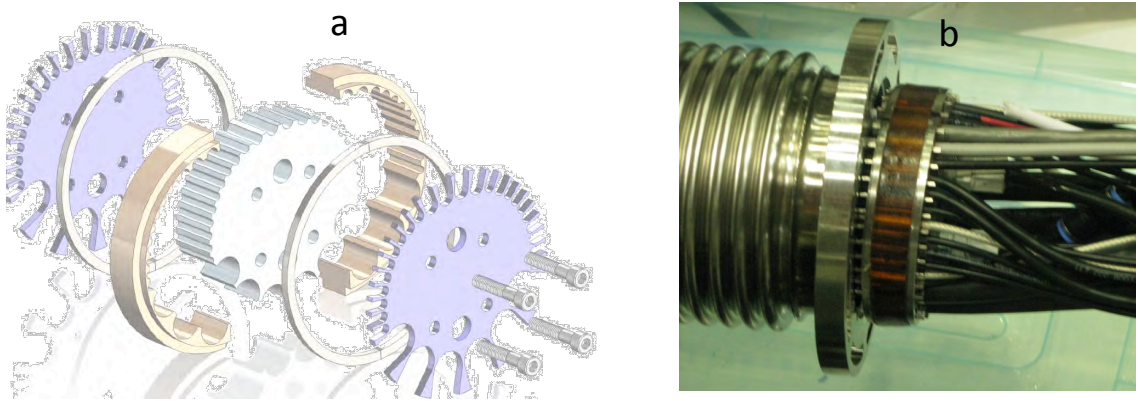}
\caption{(a) Exploded drawing of the electrical dry pipe seal pieces. (b) Electrical dry pipe seal ready to be pushed into the flange on the left.  }\label{fig:electrical_seal}
\end{figure}

To prevent humidity from entering the dry pipe a gas-tight rubber seal around the exiting electrical cables is used to seal off this volume. The parts of the seal design are shown in Fig.~\ref{fig:electrical_seal}a. A central cast rubber core (white) contains two gas line connections. Cutouts to fit the different size cables are arranged around the periphery.  An outer ring (beige) with cutouts on the inside radius fits snugly around the core and cables.  Stainless steel retaining rings hold the ring in place. The entire assembly is held together by 2 stainless steel comb plates (blue). The plug and cables are assembled together as in Fig.~\ref{fig:electrical_seal}b as a unit before being pushed into the end of the electrical dry pipe. The bolts are tightened to compress the rubber pieces sealing around the cables. This design can hold up to 3 psi pressure. During leak checks the entire electrical dry pipe volume is filled with freon at 2.5 psi and all the bellow joints are checked with freon refrigerant sniffers.  The purge gas supply runs with the electrical cables to the distribution box. The return gas is collected at the second gas fitting  in the seal and connected to a 3/8 inch exhaust line connected to the gas rack. 

\subsection{Gas Dry Pipe and Bellows}

\begin{figure}
\hfil
\includegraphics[clip=true, trim=15  28mm -15mm 40mm,width=6.5in]{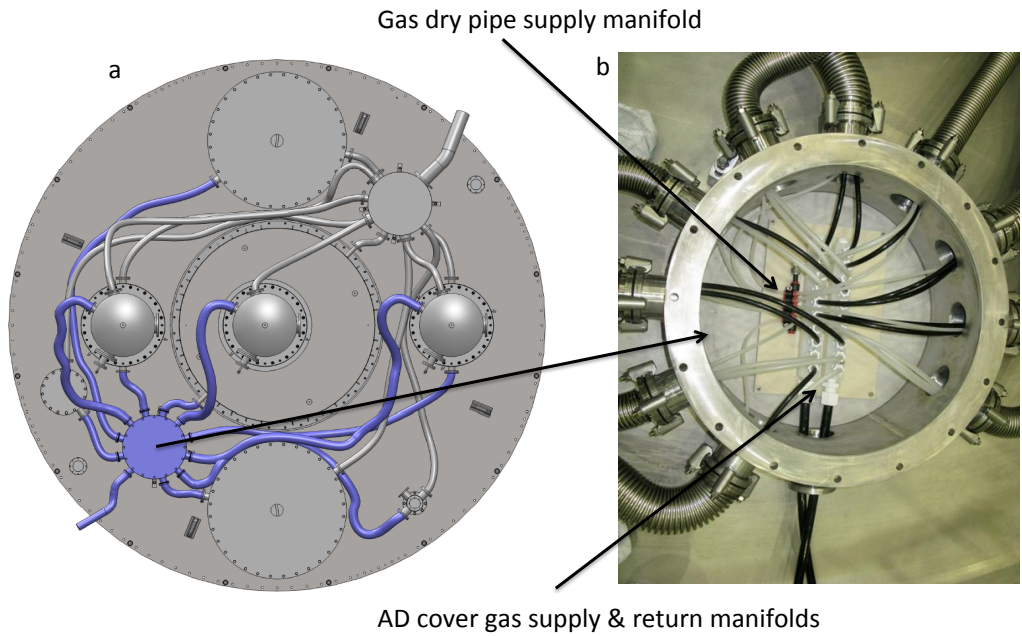}
\caption{(a) Common gas volume
(blue) of the gas distribution box and associated bellows. (b) Gas distribution box with cover gas manifolds, gas dry pipe supply manifold and tubing inside.  }\label{fig:gas}
\end{figure}

 The gas dry pipe and bellows are similar in concept to the electrical dry pipe and bellows. Cover gas flows
are gathered together in the gas distribution box (Fig.~\ref{fig:gas}b) before being routed to the surface
though a large bellows. The cover gas lines run through bellows on the lid which are isolated
from the cover gas volumes. The on-lid bellows, gas distribution box, and long gas dry pipe to
the surface comprise a single gas volume shown in Fig.~\ref{fig:gas}a. This volume is purged 
by a single 3/8 inch line which is
split into 10 lines. These smaller lines run to the end of each of the on-lid bellows. The return
gas flows back through the on-lid bellows, through the distribution box, and through the long gas supply dry
pipe to the surface. The upper end of the gas dry pipe terminates in a gas tight seal with a 3/8 inch exhaust line
connected to the gas rack.  The gas dry pipe volume and connections are also leak checked with freon.
As in the cover gas circuit, supply and return lines in the trenches outside the water pool are 12 mm copper for robustness.

\subsection{PMT Cable Bellows} 

\begin{figure}\hfil
\includegraphics[clip=true, trim=15  28mm -15mm 25mm,width=6.5in]{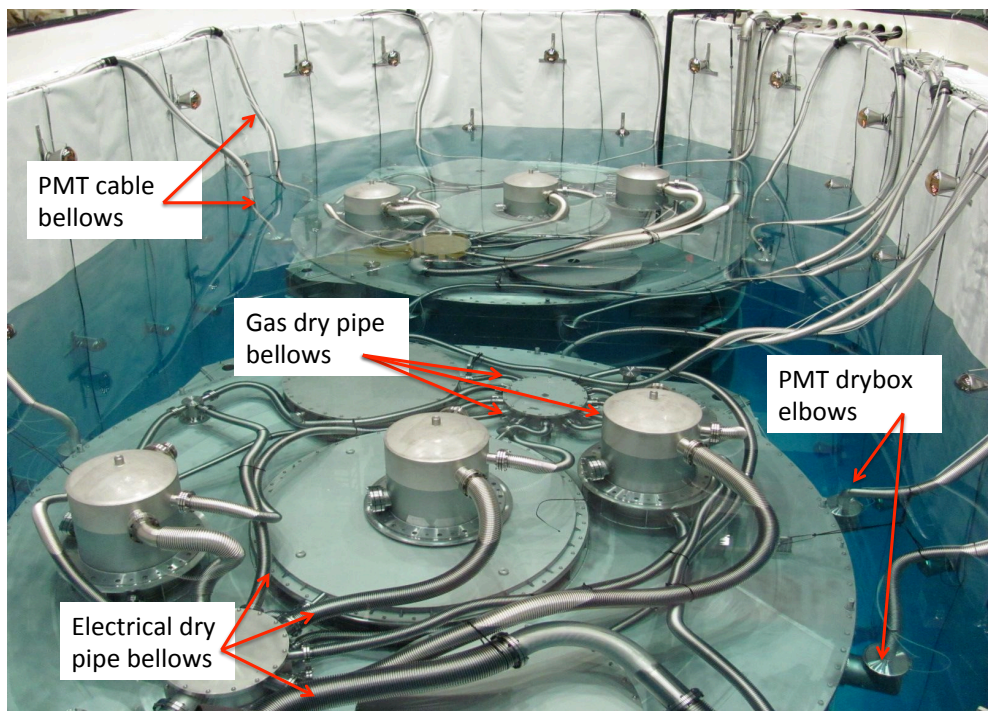}\hfil%
\caption{View of two ADs in a partially filled water pool. The 8 PMT cable bellows per AD run from the dry box elbows to the edge of the pool. 
}\label{fig:PMT}
\end{figure}

Electrical cables from the AD PMTs leave the MO volume through  eight feed-through flanges
mounted on the side wall of the AD. Connections between the PMT cables and the long cables
from the data acquisition racks  are made in large vacuum elbows (dry box elbow) attached to the flanges. The long
cables are protected from the water pool by smaller diameter bellows until they are above the water line. 
  Fig.~\ref{fig:PMT}. 
shows the dry box elbows on the side of the AD and the PMT cable bellows in a partially filled water pool.

\begin{figure}\hfil
\includegraphics[clip=true, trim=25  50mm -15mm 45mm,width=6.5in]{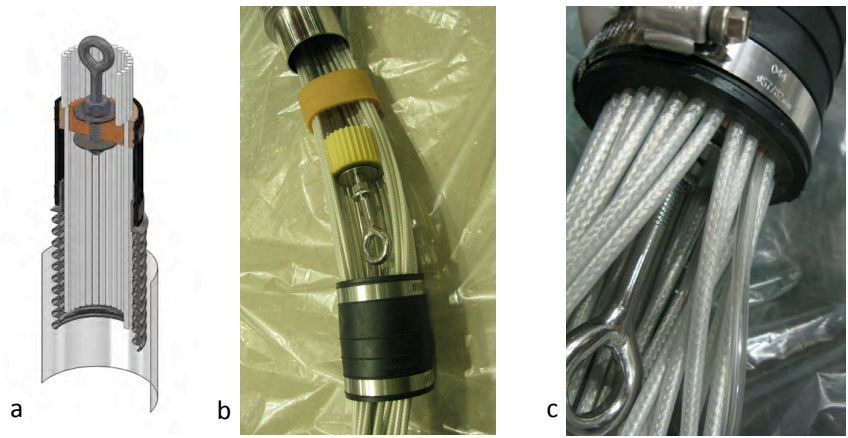}\hfil%
\caption{(a) Drawing of the bellows plug which seals around the exiting PMT cables. (b) Components of the two part plug sealing the upper end of the PMT cable bellows. (c) A completed cable assembly.}\label{fig:PMT2}
\end{figure}

A two part flexible rubber plug at
the top of the bellows (Fig.~\ref{fig:PMT2}) makes a gas-tight seal around the cables. The enclosed
volume is purged by a supply line on the drybox elbow. The return is provide by a 1/4 inch  line
through the top sealing plug. A third line at the bottom of the dry box elbow is usually closed but can be used
to test for liquid inside the elbow. Each dry box, connected bellow and gas lines  are isolated gas volumes 
under the water but are joined together above water by manifolds and run in parallel.    As with the electrical dry pipe plug the PMT cable plugs enable leak checks of the bellows connections with pressurized freon.
Unlike the other gas circuits  which typically have 1-2 cm of oil in the exit bubblers to provide back pressure, the exit bubbler
for the PMT bellows circuit is 70 cm deep, maintaining a pressure of 50 cm of water. 
This positive pressure reduces the  likelihood  of oil leaks through the PMT cables on the MO side.

\section{Gas Supply, Control and Monitoring Rack}

\begin{figure}[b]\hfil
\includegraphics[clip=true, trim=25  38mm -15mm 35mm,width=4.5in]{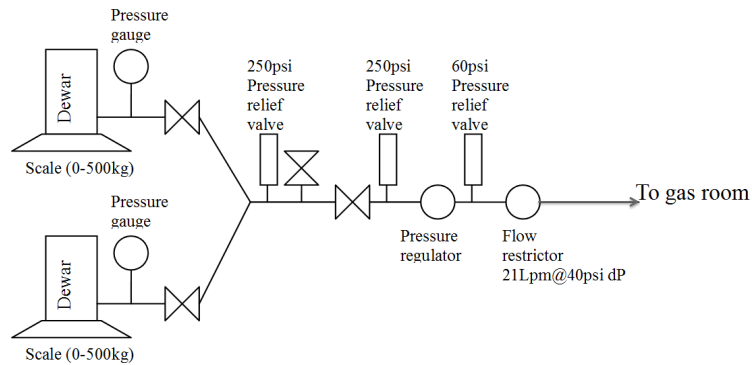}\hfil%
\caption{Schematic of the AD gas system nitrogen supply for an experimental hall. 
}\label{fig:Dewar}
\end{figure}

The gas supply system for each of the experimental halls is composed of cryogenic liquid
nitrogen dewars and a gas rack as shown in Fig. ~\ref{fig:Dewar}. Two nitrogen dewars are for safety reasons located in the tunnel
area which has a constant flesh air flow. The dewars are chained to a fixed metallic structure. 
Their weight is monitored by electrical scales. Valves allow either dewar to be connected to the supply manifold thereby ensuring uninterrupted gas flow during dewar replacements. The nitrogen is piped via flexible copper
tubing to a gas rack inside the EH gas room. As a safety measure the maximum gas flow
into gas room is limited by a flow restrictor, pressure regulator and several mechanical pressure
relief valves. Oxygen deficiency calculations were performed to determine that any potential leak would not
pose a safety and health hazard to workers underground.

\begin{figure}\hfil
\includegraphics[clip=true, trim=25  28mm -15mm 30mm,width=6.5in]{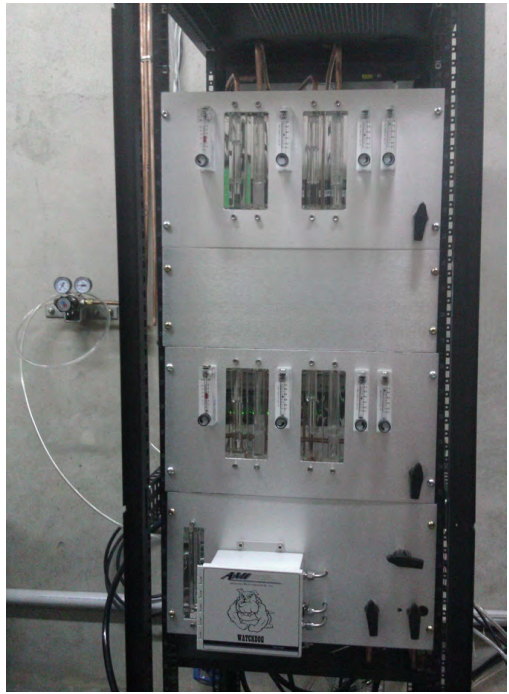}\hfil%
\caption{A gas rack for two ADs. }\label{fig:Gasrack}
\end{figure}

Inside the gas room a secondary pressure regulator reduces the pressure further before the gas
goes to the AD gas rack. A pressure transducer monitors status of the supply gas. The gas rack
is mounted with gas control elements including valves, flow meters and mineral oil filled
pressure relief bubblers, in order to adjust the gas flow. It is also equipped with humidity
sensors, an oxygen analyzer and exhaust bubblers to monitor the humidity and oxygen level of
return gas flows. All returning gas lines are joined and connected to the ventilation system.
A photograph of a near hall gas rack is shown in Fig.~\ref{fig:Gasrack}. The secondary pressure 
regulator can be seen on the wall. There is one flowmeter/bubbler panel for each AD. 
The bottom panel contains the oxygen analyzer which can be connected to either the input 
or to the returning cover gas flows.

\subsection{Flow control} 
\begin{figure}\hfil
\includegraphics[clip=true, trim=25  28mm -15mm 30mm,width=7.0in]{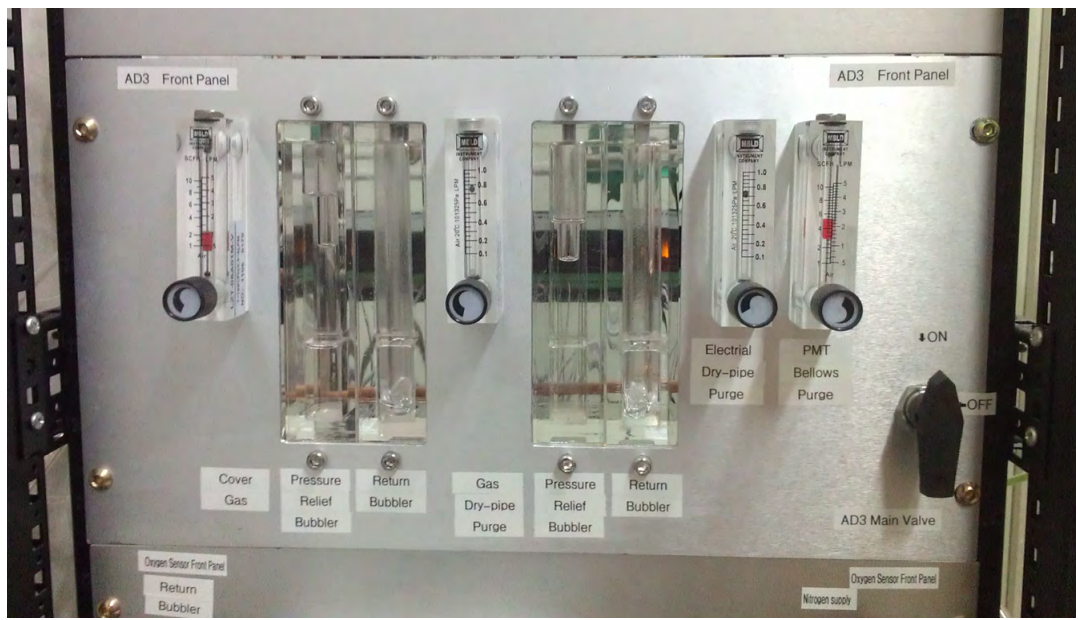}\hfil%
\caption{Front panel of the gas rack for one AD. }\label{fig:Front_panel}

\end{figure}

Each AD panel is assembled from four commercial flow meters and four custom gas bubblers as seen in Fig.~\ref{fig:Front_panel}. 
 The cover gas and gas 
dry pipe circuits have pressure relief bubblers to ensure that the maximum pressure above the liquids
is less than $\approx ~4$ cm of water.  Return bubblers keep the cover gas and gas 
dry pipe circuits  at a positive $2$ cm of water pressure above ambient.
There are two different ranges of flow meters in use. The  larger range flow meter  (0.5-5
liter per minute (lpm)) is used for the cover gas purge line. The smaller range flow meters ( 0.1-1 lpm)
are used in the gas dry pipe, electrical dry pipe,  and  PMT cable bellow lines. 

\subsection{Gas monitoring} 
Each of the gas return flows is routed past a relative humidity sensor [Honeywell HIH-3160]. A three-point calibration is done to give two linear conversions from signal to humidity value.
The zero point is obtained with supply nitrogen, while the 11.3\% and 32.7\% points are
respectively calibrated using saturated LiCl and MgCl$_2$ solution. 

A commercial oxygen analyzer, AMI Watchdog ~\cite{Watchdog}, is used to detect the oxygen level of
returning gas from the cover gas lines to the ppm level precision.
Owing to its very good linearity from ppm level to air level, a one-point calibration is done to
the analyzer using air as a 20.9\% oxygen source. The Watchdog measures less than 1~ppm of oxygen in
the supply nitrogen, using the so-calibrated analyzer.

\section{Performance}

The gas system pressures, relative humidities, and oxygen readings are recorded into the detector control system data base
and can be plotted versus time to identify long term trends.
As expected the readings fluctuate during gas outages or system maintenance but show an overall decline
in the RH and O$_2$ levels. Radon measurements require specialized equipment   and are made 
infrequently during the year. The performance history shown below are for the period of six AD running
which preceded the recent installation of the final two ADs.

\subsection{Humidity}

\begin{figure}\hfil
\includegraphics[clip=true, trim=15 15mm -15mm 25mm,width=6.0in]{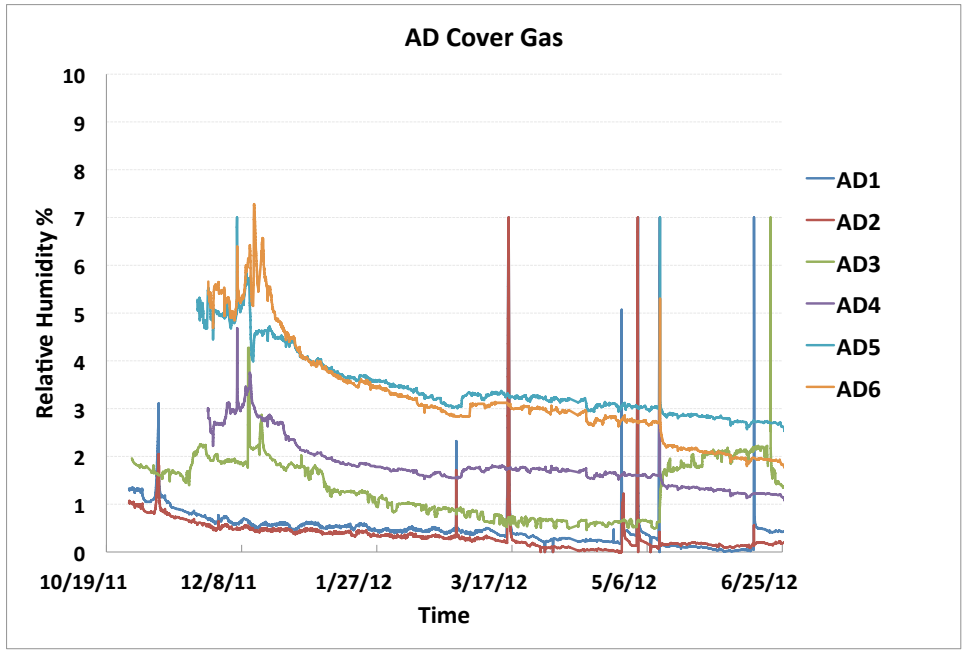}\hfil%
\caption{Relative humidity time history  of the AD cover gas returns.}\label{fig:Cover_RH}
\end{figure}

The relative humidities of the return gases are measured to provide early indications of water leaks. Tests confirm that a few cc's of water in a closed gas volume are enough to saturate the RH meter. Since the gas flows are split into eight or ten streams, a leak in only one volume should show up as a 10-12\% increase in the RH.
The  relative humidity of the cover gas returns are plotted in Fig.~\ref{fig:Cover_RH} for the past eight months.
RH is generally decreasing with time. However during gas flow outages the readings  spike (truncated to 7\%)
as stale gas in the RH meter is contaminated by small leaks in the piping and connectors. Readings recover from these outages in typically $\leq~0.5$ hours indicating that the RH of the gas over the AD is still very low since any real
RH change would have the time constant of days at two volume exchanges per day. In several ADs the RH readings are sensitive to small changes in the flow rate.  This is probably due to small leaks in the plumbing near the sensors.
Further leak checking during the next long access will try to reduce these leaks.

\begin{figure}\hfil
\includegraphics[clip=true, trim=15  15mm -15mm 25mm,width=6.0in]{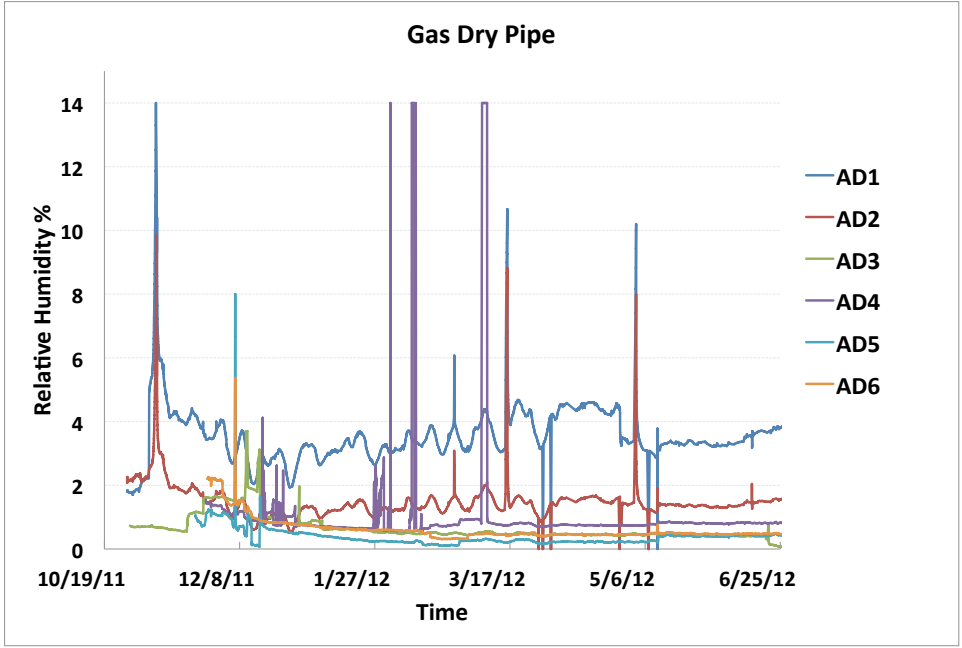}\hfil%
\caption{Relative humidity time history  of the gas dry pipe returns.}\label{fig:Gas_RH}
\end{figure}

The relative humidities of the gas dry pipe returns are plotted in Fig.~\ref{fig:Gas_RH}.
Transient spikes are  truncated  to 14\%. The RH declines for the first few months and is roughly
constant thereafter. The correlated oscillations in AD1 and AD2 are not understood but are probably
related to small oscillations in the supply pressure. A water leak in one of the 10 gas pipe bellows on  a detector
would show up as a $\geq~10\%$ step in the return RH.

\begin{figure}[b]\hfil
\includegraphics[clip=true, trim=15  15mm -15mm 25mm,width=6.0in]{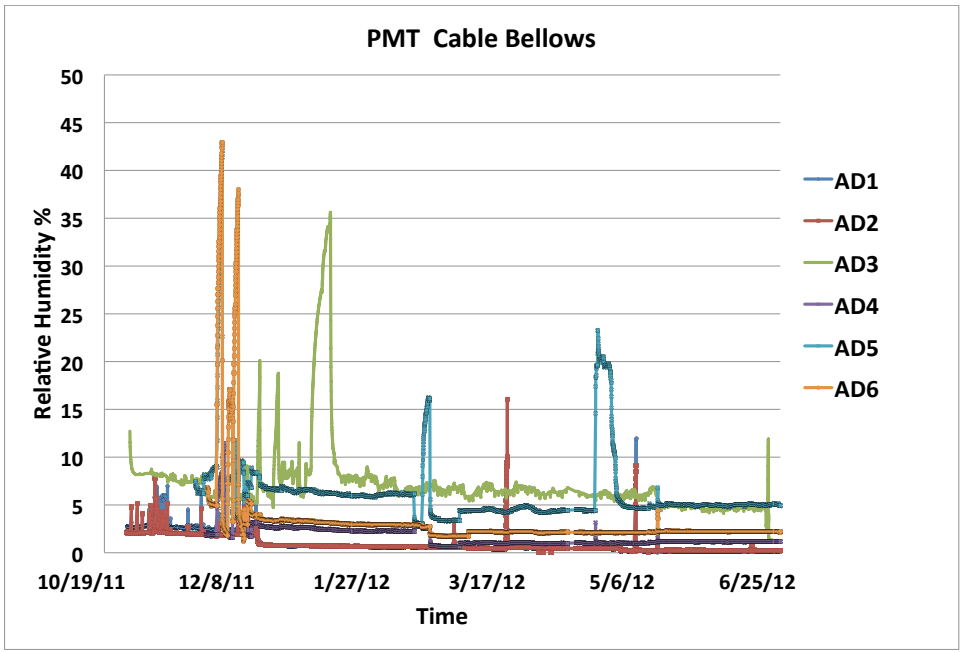}\hfil%
\caption{Relative humidity time history  of the PMT cable bellow returns.}\label{fig:PMT_v2}
\end{figure}

The  relative humidities of the PMT cable bellows returns are plotted in Fig.~\ref{fig:PMT_v2}.
Transient spikes are  truncated  to 14\%. Several high humidity periods are seen when the gas manifolds
were opened and exposed to the EH ambient humidity of $\approx~50\%$.  A water leak in one of the 8 PMT bellows on a detector
would show up as a $\geq~12\%$ step in the return RH.

\subsection{Oxygen}

The single oxygen sensor in each rack can be connected to either the input gas supply or to any of the AD cover gas returns. Measurements of the AD cover gas returns made in June and July of 2012 are shown in Table~\ref{oxygen}.
The oxygen levels were between 60 to 110 ppm.  The variation in the oxygen levels of different ADs is not understood. Sensor calibrations will be double-checked at the next opportunity. However it is more likely that the observed differences are due to small leaks within the gas rack. This is supported by data showing the return oxygen level is correlated  to the cover gas flow rate. Long term continuous measurements of AD1 over the previous seven months show a steady decline in the oxygen levels except for brief spikes associated with interruptions of the  gas flow.

\begin{table}[htdp]
\caption{Oxygen Measurements }
\begin{center}

\begin{tabular}{|c|c|}
\hline
AD & Oxygen content (ppm) \\
\hline
1 & $106.7 \pm~1.1$   \\
\hline
2 & $95.1 \pm~2.0$   \\
\hline
3 & $74.2 \pm~1.2$   \\
\hline	
4 & $83.0 \pm~1.0$   \\
  \hline
5 & $109.7 \pm~0.4$   \\
\hline
6 & $58.8\pm~0.5$   \\
\hline

\end{tabular}
\end{center}
\label{oxygen}
\end{table}

\subsection{Radon}

The radon content of the air in the experimental halls and the cover gas return flows were measured six months after the start of EH3 operation.
Since the radon levels in the return cover gas were much lower than the ambient air in the hall, particular care had to be taken in the cover gas measurements.  A custom radon detector~\cite{Lin} was developed to sample the low radon activity in the return gas with a sensitivity three times better than the  Durridge~RAD7~\cite{Durridge} detector which had been used for earlier measurements. The detector,  based on alpha spectroscopy, has the active elements enclosed within a sealed acrylic chamber which is filled by the gas leaving the detector. This gas buffer, maintained at +2 cm of water column pressure by an exit bubbler, reduces backgrounds from  the  air outside the detector so that upper limits of a few Bq/m$^3$ at the 90\% C.L. can be measured in 1/2 hour.  The detector  is automated and monitors the radon level every five minutes.  The data reported here in Table~\ref{radon} are based on integrated rates over  a > 300 min. time interval.

The experimental hall radon levels are sampled 2-3 m above the floor next to the covered water  pool. Air is drawn by a pump at 1 lpm through a dehumidifier into the detector. Neither the pump or dehumidifier are needed for the cover gas measurements as the radon detector is inserted directly into the 
return gas stream.  The ambient radon levels varied by nearly a factor of two between the experimental halls possibly due to better ventilation in the nearest hall EH1. Radons levels in the return cover gas were so low that only 90\% C.L. upper limits could be set at $\approx 0.5$ Bq/m$^3$ or better.

\begin{table}[htdp]
\caption{Radon Measurements  }
\begin{center}

\begin{tabular}{|c|c|c|c|c|c|c|}
\hline
Experimental Hall & \multicolumn{2}{|c|}{EH1} & EH2 &  \multicolumn{3}{|c|}{EH3} \\
\hline
Ambient radon (Bq/m$^3$)& \multicolumn{2}{|c|}{$136 \pm~44$}  & $221  \pm~40$ & \multicolumn{3}{|c|}{$260 \pm~40$ } \\
\hline
Detector & AD1  & AD2  & AD3 & AD4 & AD5 & AD6  \\
\hline
Measurement date 2012   & Jun 26 & Jun 27 & Jun 20 & Jun 27&Jun 28 & Jun 28\\
\hline	
Integrated time (min.)     &  370& 890&905 & 905 & 530 & 895 \\
  \hline
Counts Po-218 & 2  & 10& 6  & 12 & 6 & 9 \\
\hline
 Counts Po-214 & $0$  &1  &8 &4 &5 &3\\
 \hline
 Radon (Bq/m$^3$)  & $< 0.34$ & $< 0.50$     & $< 0.38$ &$< 0.22$  & $< 0.55$ & $< 0.45$ \\
 90\% C.L &   &  & & & &\\
\hline
\end{tabular}
\end{center}
\label{radon} 
\end{table}

\section{Summary}

The gas system for the Daya Bay antineutrino detectors has performed well. Radon levels 
in the regions above the detector liquids are over 400 times lower than the ambient environment. Oxygen levels are $\leq 110 $ ppm and continue to decline. Relative humidity measurements are also decreasing with  time and have more than the needed sensitivity to detect any possible water leaks in the future. The system is generally stable and requires infrequent maintenance.  The plug seals developed for the end of
the electrical, gas, and PMT cable bellows have proven to be indispensable in leak checking the AD gas system.

\acknowledgments

We would like to thank Dr. Kam-Biu Luk for his useful advice on setting up the gas rack and monitoring system, and also thank Dr. Yanchang Lin, Ms. Amy Leung and Prof. J. Leung  for their careful measurements of the radon levels in the experimental halls. We also thank Darrel Hamilton, Jonathan Heise, Amy Pagac and Dan Wenman of the Wisconsin Physical Sciences Laboratory for their help  installing and documenting the gas system.
This work was supported in part by the DOE Office of Science, High Energy Physics, under contract DE-FG02-95ER40896, the University of Wisconsin, the Alfred P. Sloan Foundation,
 the Research Grants Council of the Hong Kong Special Administrative Region of China (Project Nos. CUHK 1/07C and CUHK3/CRF/10), and the focused investment scheme of CUHK        

\end{document}